\documentclass[aps,prd,twocolumn,superscriptaddress,nofootinbib,floatfix]{revtex4-2}

\usepackage{amsmath,amssymb}
\usepackage{bm}
\usepackage{graphicx}
\usepackage{xcolor}
\usepackage{booktabs}
\usepackage[colorlinks=true,citecolor=blue,linkcolor=blue,urlcolor=blue]{hyperref}

\begin{document}

\title{Probing Dark Photons from Nuclear De-excitation in Reactor Neutrino Experiment}

\author{Yuanchao Lou}
\email{yuanchao_lou@nnu.edu.cn}
\affiliation{Department of Physics and Institute of Theoretical Physics, Nanjing Normal University, Nanjing, 210023, China}

\author{Lei Wu}
\email{leiwu@njnu.edu.cn}
\affiliation{Department of Physics and Institute of Theoretical Physics, Nanjing Normal University, Nanjing, 210023, China}
\affiliation{Nanjing Key Laboratory of Particle Physics and Astrophysics, Nanjing, 210023, China}

\begin{abstract}
Reactor neutrino experiments serve as powerful probes of light new physics. We investigate MeV-scale visible dark photons ($A'$) produced in nuclear reactors through nuclear de-excitation following neutron capture $N^*\to N A'$. Compared with the conventional Compton-like production process $\gamma e^-\to A'e^-$, the nuclear de-excitation yields on-shell dark photons with masses up to the nuclear transition energy. Using data from the TEXONO CsI(Tl) detector, we derive the new constraints on the kinetic mixing parameter $\epsilon$ for dark photon masses in the range $0.1\,\mathrm{MeV} < m_{A'} < 6.9\,\mathrm{MeV}$. We find that nuclear de-excitation not only extends the mass reach of reactor searches to higher dark photon masses but also provides a stronger limit than the Compton-like production process. 
\end{abstract}

\maketitle
\section{Introduction}
\label{sec:intro}

Despite compelling gravitational evidence for dark matter, its particle 
identity remains unknown~\cite{Planck:2018vyg,Bertone:2004pz}. Recent years have seen growing interest in sub-GeV ``light'' dark matter. Thermal freeze-out in this mass range generally requires new light mediators, with the dark photon being a leading candidate~\cite{Holdom:1985ag,Okun:1982xi,Arkani-Hamed:2008hhe,Bjorken:2009mm,Jaeckel:2010ni,Essig:2013lka,Fabbrichesi:2020wbt,Caputo:2021eaa,Alexander:2016aln}. Arising from a $U(1)_D$ gauge symmetry, the dark photon can mix with the Standard Model photon at the $\sim  10^{-3}$ to $10^{-6}$ level, providing a minimal mechanism for efficient annihilation. These considerations strongly motivate dedicated low-energy probes of dark photons. Such a light vector mediator can connect ordinary matter to 
MeV-scale dark-sector states and can appear in viable freeze-out 
or freeze-in dark matter scenarios~\cite{Boehm:2003hm,Feng:2008ya,
Compagnin:2022qcr,Nomura:2024jea,Knapen_2017}.

Fixed-target experiments have constrained dark photons through either
visible-decay signatures, including Orsay, KEK, E137, E141, E774,
APEX, and HPS~\cite{Davier:1989wz,Konaka:1986cb,PhysRevD.38.3375,
Riordan:1987aw,PhysRevLett.67.2942,Marsicano:2018krp,
Batell:2014mga,APEX:2011dww,Moreno:2018tlx}, or missing-energy
signatures, such as NA64~\cite{Banerjee:2019pds,NA64:2016oww},
with related searches also carried out at fixed-target configurations
and Belle II~\cite{Ge:2025aui,Liang:2022pul}.
These searches exclude
kinetic mixings roughly in the range $\epsilon\sim
10^{-5}\text{--}10^{-3}$ for MeV-to-GeV dark photons depending on
the experimental channel. Collider searches, including BaBar and KLOE-2, probe dark photons 
produced in $e^+e^-$ collisions and set limits on $\epsilon$ for 
similar mass ranges~\cite{BaBar:2014zli,KLOE-2:2011hhj,
KLOE-2:2012lii,KLOE-2:2016ydq}, while long-lived dark photon 
signatures at the LHC have also been explored~\cite{Du:2019mlc}, and accelerator-based dark photon searches
have been reviewed in~\cite{Graham:2021ggy}. Low-threshold direct detection experiments provide constraints on 
light dark photons at sub-MeV masses~\cite{An:2013yua,PandaX:2024sds}. Astrophysical and cosmological probes provide additional constraints from stellar cooling, supernova emission, BBN, and $\Delta N_{\rm eff}$~\cite{An:2013yfc,Redondo:2013lna,Chang:2016ntp,Rrapaj:2015wgs,Hardy:2016kme,Ibe:2019gpv,Arias:2012az,Fradette:2018hhl,Hufnagel:2018bjp}. 

Nuclear reactors offer an alternative laboratory source of MeV-scale
weakly coupled particles through their intense photon environment, and
have recently been explored for axion-like particle searches in
near-reactor detectors~\cite{Dai:2025kai,Gong:2026dte,Shen:2026kzy}
and for boosted dark matter via reactor neutron
scattering~\cite{Ema:2024oce}. The conventional Compton-like
production process $\gamma e^-\to A'e^-$ in reactor materials produces a flux
of dark photons detectable in nearby reactor neutrino detectors,
yielding constraints on $\epsilon$ for
$m_{A'}\lesssim 1~{\rm MeV}$~\cite{Park:2017prx}. A subsequent
reanalysis with the proper treatment of photon--hidden-photon
oscillations in the reactor medium was discussed in Ref.~\cite{Danilov:2018bks}. However, the dark photon produced from such a channel is limited by two-body kinematics, which
requires $m_{A'}\le\sqrt{m_e^2+2m_eE_\gamma}-m_e$, such as
$m_{A'}^{\rm max}\sim 3.0~{\rm MeV}$ for a typical reactor
photon energy $E_\gamma=12~{\rm MeV}$. The rapid high-energy falloff
of the reactor photon spectrum~\cite{bechteler_faissner_yogeshwar_seyfarth_1984},
$d\dot N_\gamma/dE_\gamma\sim\exp(-1.1\,E_\gamma/{\rm MeV})$
further limits the practical reach.

In this work, we investigate the visible dark photons produced in nuclear reactors through the neutron-capture nuclear de-excitation. Different from the continuum emission of Compton-like processes, this mechanism provides a discrete line source of dark photons. When a nucleus $N$ absorbs a thermal neutron and reaches an excited state $N^*$, it can de-excite either by emitting an
ordinary photon or, via kinetic mixing, a dark photon $A'$~\cite{gao2025constraintsmillichargedparticlesnuclear}. We note that the emitted dark photon carries energy $E_{A'}\simeq\omega_i$, set by the nuclear transition energy, allowing on-shell production of dark photon whenever $m_{A'}<\omega_i$. Consequently, nuclear de-excitation lifts the kinematic limit of Compton-like production, allowing reactor experiments to probe dark photons up to the nuclear transition energy. Here, we consider the representative electric dipole transition (E1) from
$^{238}{\rm U}(n,\gamma){}^{239}{\rm U}$ and
$^{10}{\rm B}(n,\gamma){}^{11}{\rm B}$ processes. For completeness, the Compton-like source is also included below its kinematic endpoint. The produced dark photons are searched for through the visible decay $A'\to e^+e^-$ and inverse Compton-like scattering $A'e^-\to\gamma e^-$ inside the detector. This paper is organized as follows. In Sec.~\ref{sec:production} we
describe the two production mechanisms and compute the 
dark photon flux at the reactor. Section~\ref{sec:detection} discusses
the visible-decay and inverse Compton-like detection channels. We
present our constraints in Sec.~\ref{sec:results} and compare with
existing bounds. Section~\ref{sec:conclusion} summarizes our
conclusions.

\section{Dark-Photon Production at Nuclear Reactors}
\label{sec:production}

The dark photon $A'$ is the massive gauge boson associated with a hidden $U(1)_D$ gauge group. Its Lagrangian, including kinetic and mass
terms as well as couplings to the SM electromagnetic current, is given by
\begin{equation}
\mathcal{L} = -\frac{1}{4} F'_{\mu\nu} F'^{\mu\nu} + \frac{1}{2} m_{A'}^2 A'_\mu A'^\mu
- \epsilon e\, A'_\mu J_{\rm EM}^\mu,
\end{equation}
where $F'_{\mu\nu} = \partial_\mu A'_\nu - \partial_\nu A'_\mu$ is the dark photon field strength tensor, \(m_{A'}\) is the dark photon mass, which may be generated, for example,
by a dark Higgs mechanism~\cite{Fabbrichesi:2020wbt} or by a
Stueckelberg mechanism~\cite{Feldman:2007wj,Feldman:2007nf}, $\epsilon$ is the kinetic-mixing parameter, and $J_{\rm EM}^\mu = \sum_f Q_f \bar{f} \gamma^\mu f$ is the SM electromagnetic current summed over all charged fermions $f$ with electric charge $Q_f$.

Neutron-capture nuclear de-excitation provides an important 
production channel of dark photon in the reactors. After neutron 
capture, an excited nuclear state $N^*$ can de-excite by 
emitting an on-shell dark photon through the kinetic mixing with the SM photon, 
\begin{equation}
n+N \to N^* \to N + A'.
\end{equation}
Since nuclear recoil corrections are negligible at MeV scales, the emitted dark photon from the $i$-th transition line carries energy $E_{A',i}\simeq\omega_i$, where $\omega_i$ is the 
corresponding transition energy, and can be produced whenever $m_{A'} < \omega_i$. The dark-photon flux at the detector from the $i$-th transition is given by
\begin{equation}
\Phi_{A',i}=\frac{\dot N_{\gamma,i}\,r_{i}}{4\pi L^2},
\label{eq:phiV}
\end{equation}
and the total de-excitation flux is
\(\Phi_{A'}^{\rm deex}=\sum_i\Phi_{A',i}\). Here $r_i$ is the dark-photon-to-photon emission ratio for the same transition line,
\begin{equation}
r_i=
\frac{\sigma(n+N \to N^* \to N + A')}{\sigma(n+N \to N^* \to N +\gamma)}.
\label{eq:ri_def}
\end{equation}
For an electric dipole transition of energy $\omega_i$, this 
ratio evaluates to~\cite{Pitrou:2019pqh}
\begin{equation}
r_{E1,i}=\epsilon^2
\left(1+\frac{m_{A'}^2}{2\omega_i^2}\right)
\sqrt{1-\frac{m_{A'}^2}{\omega_i^2}}\,
\Theta(\omega_i-m_{A'}).
\label{eq:rE1}
\end{equation}

We consider representative E1 transitions from 
$^{238}{\rm U}(n,\gamma){}^{239}{\rm U}$ and 
$^{10}{\rm B}(n,\gamma){}^{11}{\rm B}$. The selected transitions 
and their energies are listed in Table~\ref{tab:transitions}. The 
highest retained transition is the $7.007~{\rm MeV}$ E1 line from 
$^{10}{\rm B}(n,\gamma){}^{11}{\rm B}$; however, the available 
phase space $\sqrt{1-m_{A'}^2/\omega_i^2}$ vanishes as 
$m_{A'}\to\omega_i$, and the dark-photon flux becomes negligibly 
small before reaching this endpoint. We therefore restrict the scan 
to $m_{A'}\le 6.9~{\rm MeV}$, just below the kinematic limit set 
by the highest transition energy. We note that the sensitivity for 
$m_{A'}$ above the uranium transition energies is substantially 
reduced by the small $^{10}{\rm B}(n,\gamma)$ branching factor, 
as discussed below.

\begin{table}[!h]
\centering
\caption{Radiative neutron-capture transitions used in this 
work, with reaction, transition energy $\omega_i$, 
multipolarity, photon intensity $I_\gamma$ per 
100 radiative captures, and radiative-capture branching 
factor $f_{n\gamma}$.}
\label{tab:transitions}
\begin{tabular}{lcccc}
\hline\hline
Reaction & $\omega_i$ [MeV] &\, Type & $I_\gamma$ & $f_{n\gamma}$ \\
\hline
$^{238}{\rm U}(n,\gamma){}^{239}{\rm U}$ & 3.297 & E1 & 13.21 & 1 \\
$^{238}{\rm U}(n,\gamma){}^{239}{\rm U}$ & 4.060 & E1 & 6.95  & 1 \\
$^{10}{\rm B}(n,\gamma){}^{11}{\rm B}$  & 4.711 & E1 & 25.58 \,
    & $7.9\times10^{-5}$ \\
$^{10}{\rm B}(n,\gamma){}^{11}{\rm B}$  & 7.007 & E1 & 55.10 \,
    & $7.9\times10^{-5}$ \\
\hline\hline
\end{tabular}
\end{table}

The ordinary photon emission rate for transition $i$ from 
capture isotope $N$ is given by
\begin{equation}
\dot N_{\gamma,i}
=
N_n\,
\frac{Y_n^{(N)}}{\sum_N Y_n^{(N)}}\,
f_{n\gamma}^{(N)}\,
\frac{I_\gamma^{(N,\omega_i)}}{100},
\label{eq:Ngamma}
\end{equation}
where $N_n$ is the total neutron production rate, $Y_n^{(N)}$ 
is the neutron-absorption yield per fission for isotope $N$, 
$I_\gamma^{(N,\omega_i)}$ is the number of photons of energy 
$\omega_i$ emitted per 100 radiative captures, and 
$f_{n\gamma}^{(N)}$ is the fraction of neutron absorptions 
proceeding via the radiative-capture channel. We take 
$f_{n\gamma}^{(^{238}{\rm U})}=1$, since radiative capture 
dominates for $^{238}{\rm U}$. For $^{10}{\rm B}$, the dominant 
absorption channel is $^{10}{\rm B}(n,\alpha){}^{7}{\rm Li}$ 
with thermal cross section $3837~{\rm b}$, while 
$\sigma[^{10}{\rm B}(n,\gamma){}^{11}{\rm B}]\simeq 
0.305~{\rm b}$, giving 
$f_{n\gamma}^{(^{10}{\rm B})}\simeq 7.9\times10^{-5}$~\cite{Mughabghab:2006}. 
The total neutron production rate $N_n$ is determined by the 
reactor thermal power. Each fission releases an average of 
$200~{\rm MeV}$ energy~\cite{schunck2024nuclear} and produces 
$2.5$ neutrons on average~\cite{TEXONO:2005fmk}, giving 
$N_n = 7.8\times10^{19} \times P$, where $P$ is the reactor 
thermal power in GW. For the TEXONO configuration with 
$P=2.9~{\rm GW}$, this yields 
$N_n\simeq 2.3\times10^{20}~{\rm s}^{-1}$. The 
neutron-absorption yields and transition intensities are taken 
from Refs.~\cite{TEXONO:2005fmk,ENSDF}.

For the nuclear de-excitation contribution, we focus on the dark photons
emitted directly in the primary transition \(N^\ast\to N A'\). Ordinary
transition photons may also propagate through reactor materials and
subsequently produce dark photons through secondary Compton-like
conversion, \(\gamma e^-\to A'e^-\). Such a component would require a
reactor-specific photon-transport treatment, including the geometry,
material composition, and attenuation history of the photons, and is
not included here.

The continuum Compton-like contribution is treated separately using the reactor photon spectrum. We adopt the analytical
parametrization obtained for the FRJ-1 reactor core, which applies for
photon energies $E_\gamma \gtrsim 200~{\rm keV}$~\cite{bechteler_faissner_yogeshwar_seyfarth_1984}.
\begin{equation}
\frac{d\dot N_\gamma}{dE_\gamma}
=
5.8\times10^{17}~\frac{1}{{\rm MeV\cdot s}}\,
\frac{P}{\rm MW}\,
e^{-1.1E_\gamma/{\rm MeV}}.
\label{eq:reactor_spectrum}
\end{equation}
This spectrum is dominated by prompt fission $\gamma$-rays and covers energies up to several MeV. The resulting dark-photon flux 
at the detector is
\begin{equation}
\frac{d\Phi_{A'}^{\rm comp}}{dE_{A'}}
=
\frac{1}{4\pi L^2}
\int dE_\gamma\,
\frac{d\dot N_\gamma}{dE_\gamma}\,
\frac{1}{\sigma_{\rm tot}(E_\gamma)}
\frac{d\sigma_{\gamma e\to A'e}}{dE_{A'}},
\label{eq:compton_flux}
\end{equation}
where $L=28~{\rm m}$ is the reactor-to-detector distance for
TEXONO~\cite{TEXONO:2009knm}. The factor
$(1/\sigma_{\rm tot})\,d\sigma_{\gamma e\to A'e}/dE_{A'}$
is the differential probability for a reactor photon interaction to
produce a dark photon with energy $E_{A'}$. We take
$\sigma_{\rm tot}(E_\gamma)
=
\sigma_{\rm SM}(E_\gamma)
+
\sigma_{\gamma e\to A'e}(E_\gamma)
\simeq
\sigma_{\rm SM}(E_\gamma)$,
since the dark-photon production cross section is suppressed by
$\epsilon^2 \ll 1$. The SM cross section
$\sigma_{\rm SM}(E_\gamma)$ is taken from the XCOM photon cross-section
database~\cite{XCOM2010}. 

\begin{figure}[t]
\centering
\includegraphics[width=\columnwidth]{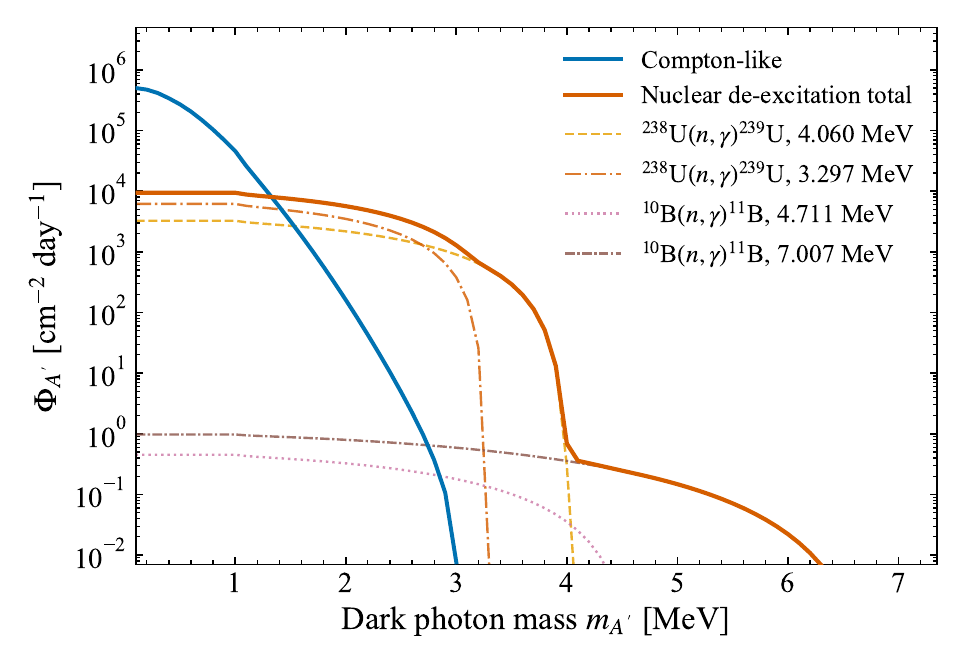}
\caption{Integrated dark-photon flux $\Phi_{A'}$ at the TEXONO 
detector as a function of dark-photon mass $m_{A'}$, for 
$\epsilon=10^{-6}$. The blue curve shows the Compton-like flux, 
obtained by integrating $d\Phi_{A'}/dE_{A'}$ over all energies, 
which closes at $m_{A'}^{\rm max}=3.03~{\rm MeV}$. The dashed 
and dash-dotted curves show individual de-excitation line 
contributions from $^{238}{\rm U}(n,\gamma){}^{239}{\rm U}$ 
(3.297 and 4.060~MeV E1) and $^{10}{\rm B}(n,\gamma){}^{11}{\rm B}$ 
(4.711 and 7.007~MeV E1), and the orange curve gives their sum. 
The de-excitation channel extends reactor sensitivity to 
$m_{A'}\lesssim 6.9~{\rm MeV}$, well beyond the Compton-like 
ceiling.}
\label{fig:deex_flux}
\end{figure}

Figure~\ref{fig:deex_flux} shows the integrated dark-photon 
flux $\Phi_{A'}$ as a function of dark-photon mass $m_{A'}$ 
for $\epsilon=10^{-6}$, where the Compton-like flux is included 
for comparison as the energy-integrated total 
$\Phi_{A'}^{\rm Compton}=\int dE_{A'}\,d\Phi_{A'}/dE_{A'}$.
The individual de-excitation lines (dashed and dash-dotted 
curves) each contribute a flat plateau that terminates sharply 
at $m_{A'}=\omega_i$, while the total de-excitation flux 
(orange) extends to $m_{A'}\lesssim 6.9~{\rm MeV}$, well 
beyond the mass endpoint of the dark photon from the Compton-like process, $m_{A'} = 3.03~{\rm MeV}$.

\section{Detection and Event Rates}
\label{sec:detection}

The TEXONO CsI(Tl) detector is located at the Kuo-Sheng Reactor Neutrino
Laboratory, $L=28~{\rm m}$ from a $P=2.9~{\rm GW}$ reactor
core~\cite{TEXONO:2009knm}. The detector has a total mass
$M_{\rm det}=187~{\rm kg}$, and we use the TEXONO analysis window
$3~{\rm MeV}<E_{\rm dep}<8~{\rm MeV}$. The number of target electrons
per unit detector mass is $N_e/M_{\rm det}=108\,N_A/M_{\rm CsI}
\simeq 2.51\times10^{26}~{\rm kg^{-1}}$, where
$M_{\rm CsI}=259.81~{\rm g\,mol^{-1}}$ and 108 is the number of
electrons per CsI formula unit.

Dark photons arriving at the detector can give two visible signatures:
decay inside the detector, $A'\to e^+e^-$, and inverse Compton-like
scattering on electrons, $A'e^-\to\gamma e^-$. For visible decays,
the partial width is
\begin{equation}
\Gamma_{A'\to e^+e^-}
=
\frac{\epsilon^2\alpha m_{A'}}{3}
\left(1+\frac{2m_e^2}{m_{A'}^2}\right)
\sqrt{1-\frac{4m_e^2}{m_{A'}^2}},
\label{eq:width_ee_visible}
\end{equation}
for $m_{A'}>2m_e$. The boosted decay length is
\begin{equation}
\ell_{A'}(E_{A'})
=
\frac{p_{A'}}{m_{A'}\,\Gamma_{A'\to e^+e^-}},
\label{eq:decay_length}
\end{equation}
where $p_{A'}=\sqrt{E_{A'}^2-m_{A'}^2}$. Numerically, $\ell_{A'} \sim 4\times10^{10}~{\rm m}\,
(\epsilon/10^{-6})^{-2}(m_{A'}/{\rm MeV})^{-2}(E_{A'}/{\rm MeV})$,
which far exceeds the baseline for the TEXONO
configuration in the coupling and mass range of interest. The survival
probability $P_{\rm surv}(L,E_{A'})=\exp[-L/\ell_{A'}(E_{A'})]$ is
close to unity throughout most of the parameter space considered here;
we nevertheless retain this factor in the numerical calculation. For
decays inside the detector, the visible energy is
$E_{\rm dep}\simeq E_{A'}$.

For inverse Compton-like scattering~\cite{Su:2021jvk}, we compute the lab-frame
differential cross section $d\sigma_{A'e\to\gamma e}/dE_r$, where
$E_r$ is the outgoing photon energy. The squared matrix element and the
kinematic transformation to the lab frame are summarized in
Appendix~\ref{app:kinematics}. Assuming full containment of the
final-state photon and electron, energy conservation gives
$E_{\rm dep}\simeq E_r+T_e=E_{A'}$, where $T_e$ is the recoil electron
kinetic energy. As a consistency check, in the limit $m_{A'}\ll m_e$
our result reduces to
\begin{equation}
\frac{d\sigma_{A'e\to\gamma e}}{dE_r}
\simeq
\frac{2}{3}\epsilon^2
\frac{d\sigma_C}{dE_r}
\left[1+\mathcal{O}\!\left(\frac{m_{A'}^2}{m_e^2}\right)\right],
\label{eq:park_limit}
\end{equation}
in agreement with the approximation used in Ref.~\cite{Park:2017prx}.

Having specified the two detection channels, we now write the corresponding
event-rate expressions for the two reactor production sources. For
the continuous Compton-like source, the scattering contribution per
unit detector mass is
\begin{equation}
\frac{dR_{\rm scat}^{\rm comp}}{dE_{A'}}
=
\frac{N_e}{M_{\rm det}}
\frac{d\Phi_{A'}^{\rm comp}}{dE_{A'}}
\sigma_{A'e\to\gamma e}(E_{A'}),
\label{eq:dRscat_comp}
\end{equation}
where
\begin{equation}
\sigma_{A'e\to\gamma e}(E_{A'})
=
\int_{E_r^{\rm min}}^{E_r^{\rm max}}
dE_r\,
\frac{d\sigma_{A'e\to\gamma e}}{dE_r}.
\end{equation}
The corresponding visible-decay contribution in the long-lived limit is
\begin{equation}
\frac{dR_{\rm dec}^{\rm comp}}{dE_{A'}}
\simeq
\frac{V_{\rm det}}{M_{\rm det}}
\frac{d\Phi_{A'}^{\rm comp}}{dE_{A'}}
\frac{P_{\rm surv}(L,E_{A'})}{\ell_{A'}(E_{A'})},
\label{eq:dRdec_comp}
\end{equation}
where $V_{\rm det}$ is the active detector volume.

For the nuclear de-excitation source, the emitted dark photons are
monoenergetic, $E_{A',i}\simeq\omega_i$. The scattering rate per unit
detector mass from line $i$ is
\begin{equation}
R_{{\rm scat},i}^{\rm deex}
=
\frac{N_e}{M_{\rm det}}
\,\Phi_{A',i}\,
\sigma_{A'e\to\gamma e}(\omega_i),
\label{eq:Rscat_deex}
\end{equation}
while the visible-decay rate is
\begin{equation}
R_{{\rm dec},i}^{\rm deex}
\simeq
\frac{V_{\rm det}}{M_{\rm det}}
\,\Phi_{A',i}\,
\frac{P_{\rm surv}(L,\omega_i)}{\ell_{A'}(\omega_i)}.
\label{eq:Rdec_deex}
\end{equation}

\begin{figure}[!t]
\centering
\includegraphics[width=\columnwidth]{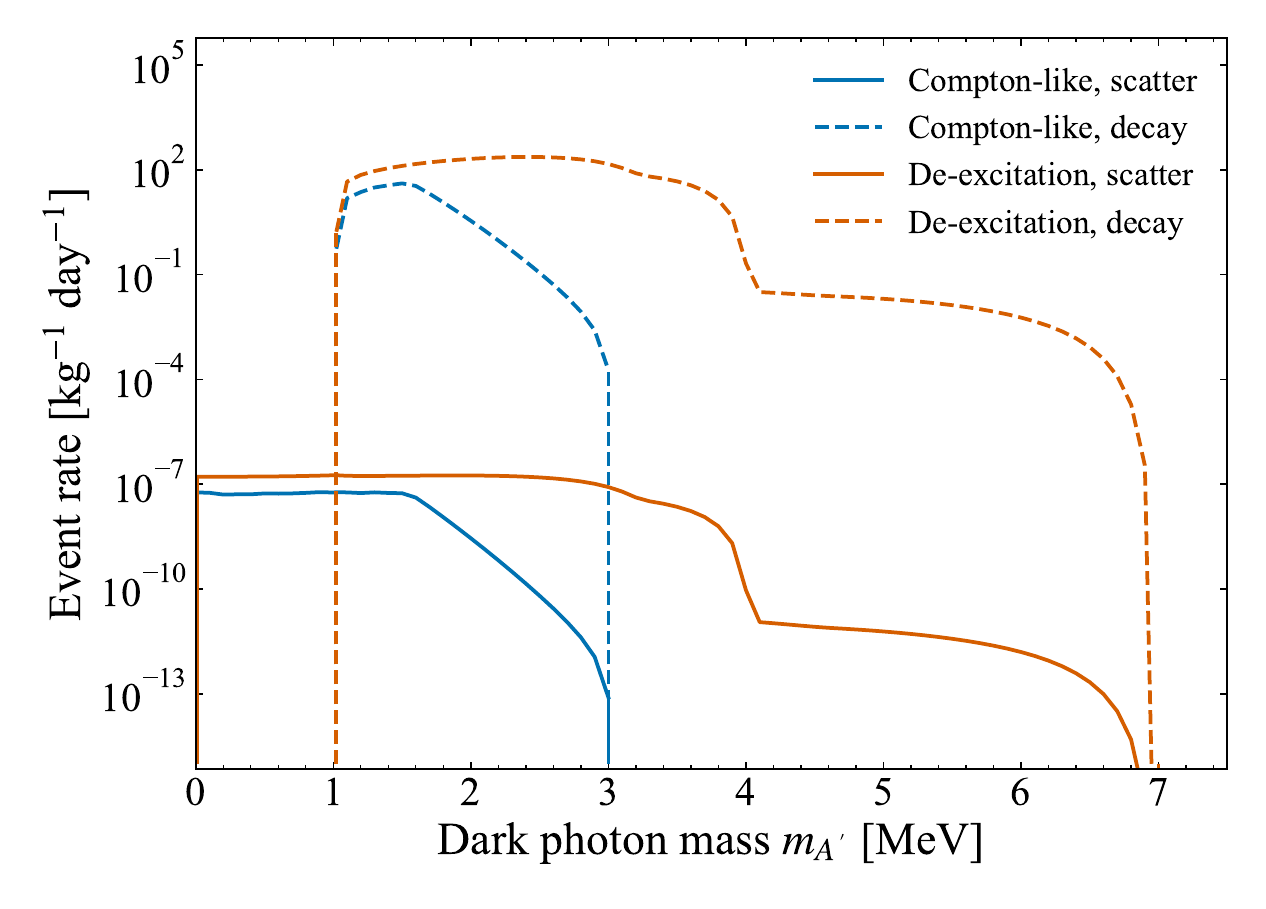}
\caption{Event-rate components at the TEXONO CsI detector for
$\epsilon=10^{-6}$. Solid curves show inverse Compton-like scattering
$A'e^-\to\gamma e^-$; dashed curves show visible decays
$A'\to e^+e^-$. Blue and orange curves correspond to Compton-like
production and nuclear de-excitation, respectively.}
\label{fig:rate_components}
\end{figure}

Figure~\ref{fig:rate_components} compares the scattering and decay
rate components for the two reactor production mechanisms. The
Compton-like contribution follows the continuum reactor photon flux,
whereas the de-excitation contribution reflects the selected nuclear
transition lines. In the parameter point shown, the visible-decay
contribution dominates once $A'\to e^+e^-$ is kinematically open.

\section{Constraints}
\label{sec:results}
We constrain the two visible signatures with different TEXONO CsI(Tl)
observables. For visible decays, $A'\to e^+e^-$, we use the TEXONO
three-hit cosmic-ray-unrelated pair-production spectrum, which selects
one primary energy deposition accompanied by two $511~{\rm keV}$
annihilation photons in neighboring crystals~\cite{TEXONO:2009knm}. For
inverse Compton-like scattering, $A'e^-\to\gamma e^-$, we instead use
the inclusive TEXONO $\bar\nu_e-e^-$ measurement in the same energy
window, requiring the predicted scattering contribution to satisfy
$N_{\rm scat}<N_{\rm scat}^{95}=1.96\times100.6\simeq197.2$ at
95\% C.L. At each mass point, the final limit on $\epsilon$ is taken
as the stronger of the two constraints; in practice, the scattering
constraint dominates below the $e^+e^-$ threshold and the
visible-decay constraint above it.

For the Compton-like source, the dark-photon spectrum is continuous 
in $E_{A'}$. The number of inverse Compton-like scattering events is 
obtained by integrating the predicted deposited-energy spectrum over 
the TEXONO analysis window,
\begin{equation}
N_{\rm scat}^{\rm comp}
=
M_{\rm det}\,T
\int_{3~{\rm MeV}}^{8~{\rm MeV}}
dE_{A'}\,
\frac{dR_{\rm scat}^{\rm comp}}{dE_{A'}},
\label{eq:Nscat_compton_texono}
\end{equation}
where $T=160~{\rm days}$. Parameter points satisfying 
$N_{\rm scat}^{\rm comp}>N_{\rm scat}^{95}$ are excluded. The same Compton-produced dark-photon flux can also give visible decays
inside the detector. This contribution is compared with the TEXONO
three-hit pair-production spectrum. The predicted binned decay rate is
\begin{equation}
\left(\frac{dR_{\rm dec}^{\rm comp}}{dE}\right)_b
=
\frac{1}{\Delta E_b}
\int_{E_b^{\rm min}}^{E_b^{\rm max}}
dE_{\rm dep}\,
\frac{dR_{\rm dec}^{\rm comp}}{dE_{\rm dep}},
\label{eq:bin_decay_compton}
\end{equation}
where $dR_{\rm dec}^{\rm comp}/dE_{\rm dep}$ is obtained from
Eq.~\eqref{eq:dRdec_comp}. We use a bin-wise $\chi^2$ statistic,
\begin{equation}
\chi^2_{\rm dec}(\epsilon)
=
\sum_b
\left[
\frac{
\max\!\left(
0,
\left(dR_{\rm dec}^{\rm comp}/dE\right)_b
-
R_b^{\rm obs}
\right)
}{
\sigma_b
}
\right]^2 ,
\label{eq:chi2_decay_compton}
\end{equation}
where $R_b^{\rm obs}$ and $\sigma_b$ are the measured central value and
experimental uncertainty in bin $b$ of the TEXONO three-hit
cosmic-ray-unrelated pair-production spectrum. The 
95\% C.L. upper limit is obtained from $\Delta\chi^2=3.84$.

For the nuclear de-excitation source, the dark photons are produced 
as monoenergetic lines with $E_{A',i}\simeq\omega_i$. The 
inverse Compton-like scattering contribution is treated with the 
inclusive TEXONO event-count constraint,
\begin{equation}
N_{\rm scat}^{\rm deex}
=
M_{\rm det}\,T
\sum_i
R_{{\rm scat},i}^{\rm deex}\,
\Theta(3~{\rm MeV}<\omega_i<8~{\rm MeV}),
\label{eq:Nscat_deex_texono}
\end{equation}
where $R_{{\rm scat},i}^{\rm deex}$ is the scattering event rate per 
unit detector mass from line $i$. Parameter points satisfying 
$N_{\rm scat}^{\rm deex}>N_{\rm scat}^{95}$ are excluded. The same de-excitation flux can also produce visible decays inside the
detector. Since this signal is line-like, we compare it with the TEXONO
three-hit pair-production spectrum using a bin-by-bin rate cap. The 
decay rate $R_{{\rm dec},i}^{\rm deex}$ for line $i$ is converted 
into a bin-averaged differential rate,
\begin{equation}
\mathcal{R}_b^{\rm deex}
=
\frac{1}{\Delta E_b}
\sum_{i\in b}
R_{{\rm dec},i}^{\rm deex},
\label{eq:deex_bin_avg_rate}
\end{equation}
where the sum runs over all de-excitation lines whose transition 
energies fall in bin $b$. We conservatively exclude a parameter 
point if $\mathcal{R}_b^{\rm deex}$ exceeds the corresponding 
TEXONO three-hit cosmic-ray-unrelated rate cap in any bin 
containing a de-excitation line, following the conservative recast 
strategy of 
Ref.~\cite{gao2025constraintsmillichargedparticlesnuclear}.

\begin{figure}[!t]
\centering
\includegraphics[width=\columnwidth]{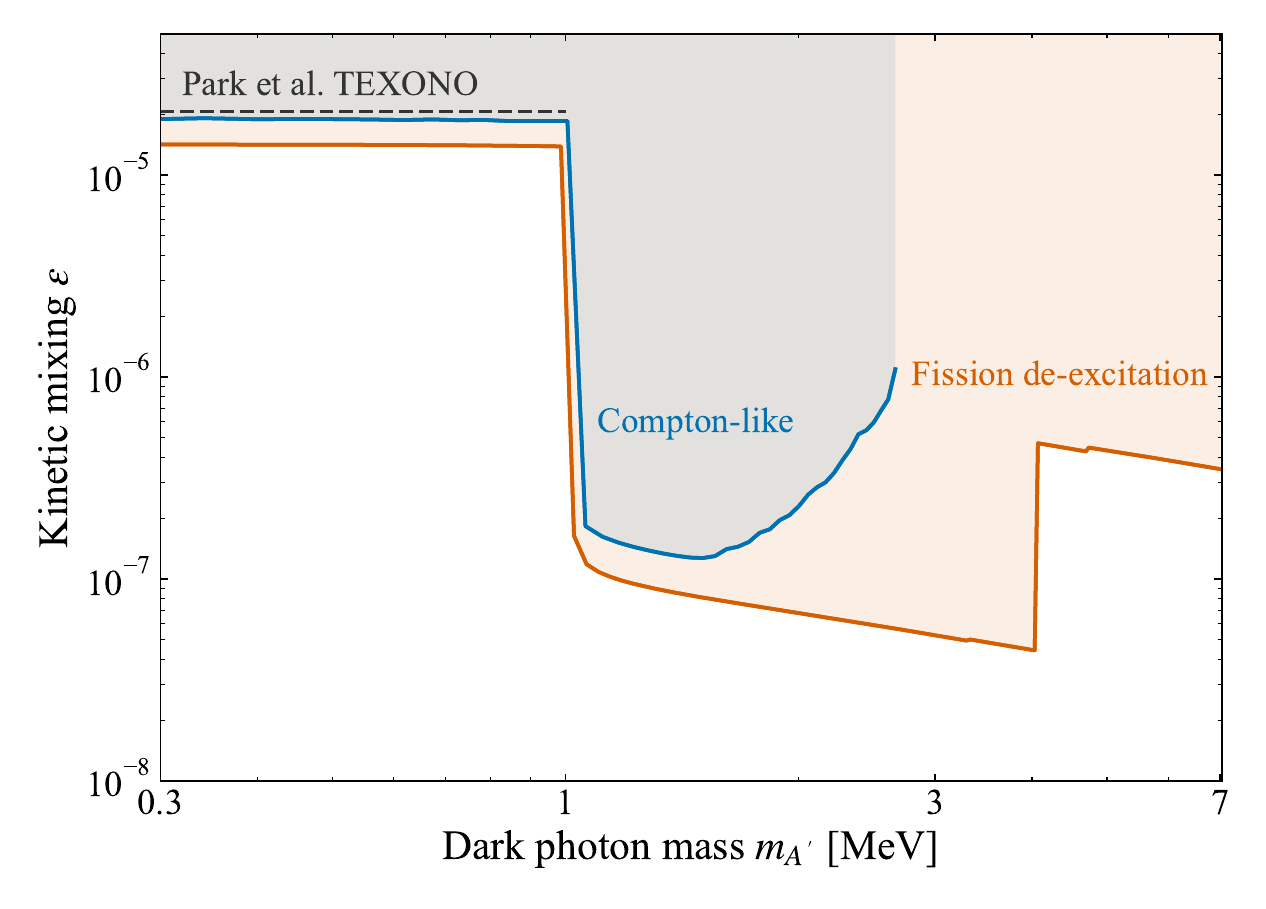}
\caption{Constraints on the kinetic mixing parameter $\epsilon$ as a
function of dark-photon mass $m_{A'}$, derived from the TEXONO CsI(Tl)
data. The blue curve shows the constraint from Compton-like reactor
production, combining the inclusive TEXONO event-count constraint for
inverse Compton-like scattering with the H3 pair-production recast for
visible decays. The orange curve shows the corresponding constraint from
neutron-capture nuclear de-excitation, where the line-like visible-decay
signal is treated with a conservative bin-cap procedure. The gray
reference line shows the previous TEXONO reactor Compton-like limit of
Ref.~\cite{Park:2017prx}.}
\label{fig:constraints}
\end{figure}

Figure~\ref{fig:constraints} shows the resulting TEXONO constraints in
the $(m_{A'},\epsilon)$ plane. For each production mechanism, the
inverse Compton-like scattering contribution is constrained using the
inclusive TEXONO $\bar\nu_e -e^-$ event count limit, while visible
decays are constrained using the TEXONO three-hit pair production data.
For the continuous Compton-like source, the visible decay contribution
is recast with an upper-only binned $\chi^2$. For the monoenergetic
nuclear de-excitation source, the line-like visible-decay contribution
is constrained with the conservative lower envelope bin cap procedure.
The two reactor sources probe qualitatively different regions of 
phase space. The Compton-like contribution is controlled by the 
high-energy tail of the reactor photon spectrum and rapidly loses 
support near its kinematic endpoint. In contrast, neutron-capture 
de-excitation provides localized dark photon sources at discrete 
nuclear transition energies. The resulting constraint therefore 
reflects a set of line-like probes, with the strongest sensitivity 
occurring when an intense transition falls inside the TEXONO 
analysis window and contributes to either the inclusive scattering 
constraint or the three-hit visible decay constraint.

It is useful to compare the reactor constraints with other probes.
Beam-dump experiments such as E137 constrain dark photons produced in
high-energy electromagnetic showers over macroscopic
baselines~\cite{Batell:2014mga,Marsicano:2018krp}; their sensitivity depends on the production rate, shielding 
survival, and detector geometry, probing a distinct regime 
from the short-baseline MeV-scale reactor source considered here. Astrophysical and
cosmological bounds from stellar cooling, supernovae, BBN, and
$\Delta N_{\rm eff}$ can be powerful, but depend on in-medium
production, trapping, and the thermal history of the dark
sector~\cite{An:2013yfc,Chang:2016ntp,Fradette:2018hhl,Ibe:2019gpv}.
The reactor constraints derived here instead are based on measured
reactor and detector inputs, without invoking astrophysical or
cosmological assumptions.

\section{Conclusion}
\label{sec:conclusion}

We have studied MeV-scale visible dark photons produced at nuclear
reactors through two mechanisms. The first is
Compton-like conversion of reactor photons on electrons,
$\gamma e^-\to A'e^-$, which produces a continuous dark photon
spectrum but is kinematically limited to $m_{A'}\lesssim 3~{\rm MeV}$
for the reactor photon energies relevant to TEXONO. The second is
neutron-capture nuclear de-excitation, $N^*\to N A'$, in which an
on-shell dark photon is emitted in place of an ordinary transition
photon. Since this process is controlled by discrete nuclear transition
energies, it can extend the reactor dark-photon source above the
Compton-like endpoint.

Using selected E1 transitions from
$^{238}{\rm U}(n,\gamma){}^{239}{\rm U}$ and
$^{10}{\rm B}(n,\gamma){}^{11}{\rm B}$, we computed the
corresponding dark-photon flux at the TEXONO detector. We considered
two visible detection signatures in the CsI(Tl) array:
inverse Compton-like scattering, $A'e^-\to\gamma e^-$, and visible
decay, $A'\to e^+e^-$. The scattering contribution was constrained
using the inclusive TEXONO $\bar\nu_e-e^-$ event-count measurement in
the $3$--$8~{\rm MeV}$ deposited-energy window, while the visible
decay contribution was constrained using the TEXONO three-hit
pair-production spectrum. This treatment separates the two
experimental topologies and allows both continuous Compton-like sources
and discrete nuclear de-excitation lines to be tested consistently.

The resulting limits constrain the kinetic mixing parameter 
$\epsilon$ in the MeV-scale mass range. Beyond extending the 
mass coverage, the key physics point is that reactor neutron 
capture provides a calibrated line source of on-shell dark 
photons, tying the reactor sensitivity to measured nuclear 
transition energies and intensities rather than to the 
exponentially suppressed high-energy tail of the continuum 
photon spectrum.

These limits provide an independent laboratory probe of MeV-scale 
visible dark photons, distinct from accelerator beam-dump, 
astrophysical, and cosmological constraints. Beam-dump experiments 
probe dark photons produced in high-energy electromagnetic showers 
and are sensitive to the interplay of production, shielding, 
propagation, and detector geometry. Astrophysical and cosmological 
bounds can be powerful, but depend on assumptions about in-medium 
production, trapping, thermal history, and available decay modes. 

The reactor constraints derived here are instead based on measured
reactor and detector inputs, without invoking astrophysical or
cosmological assumptions. Further improvements can come from a more
complete set of neutron-capture transitions and from applying the same
strategy to high-resolution near-reactor detectors, such as
TAO~\cite{JUNO:2020ijm}, which could strengthen the sensitivity to
line-like de-excitation signals. The same production formalism can also
be adapted to invisible dark-photon decays sourcing  light dark
matter via reactor de-excitation.

\begin{acknowledgments}
We thank Maxim Pospelov for helpful discussions. L.W. is supported by 
the NNSFC under Grant No.~12275134 and No.~12335005.
\end{acknowledgments}

\appendix

\section{Kinematics for the Inverse Compton-like Process}
\label{app:kinematics}

In this appendix we summarize the kinematic expressions used to compute
the inverse Compton-like detection process,
\begin{equation}
A'(p_1) + e^-(p_2) \to \gamma(k_1) + e^-(k_2),
\end{equation}
with the initial electron at rest in the laboratory frame. The Mandelstam
variable is
\begin{equation}
s
=
m_e^2 + m_{A'}^2 + 2m_e E_{A'},
\end{equation}
where $E_{A'}$ is the lab-frame dark-photon energy. We define the
dimensionless ratios
\begin{equation}
x_e \equiv \frac{m_e}{\sqrt{s}},
\qquad
x_{A'} \equiv \frac{m_{A'}}{\sqrt{s}},
\end{equation}
and
\begin{equation}
\kappa
=
\frac{1}{2}
\sqrt{
\left(1+x_e^2-x_{A'}^2\right)^2
-4x_e^2
}.
\end{equation}
The center-of-mass energy of the outgoing photon is
\begin{equation}
E_\gamma^\ast
=
\frac{s-m_e^2}{2\sqrt{s}}.
\end{equation}

After averaging over initial-state polarizations and summing over
final-state polarizations, the COM-frame differential cross section can
be written as
\begin{equation}
\frac{d\sigma_{A'e\to\gamma e}}{d\Omega^\ast}
=
\frac{\alpha^2\epsilon^2}{48s}
\frac{1-x_e^2}{\kappa}
\left(T_1+T_2+T_3\right),
\label{eq:inverse_domega_appendix}
\end{equation}
where
\begin{align}
T_1
&=
\frac{16}{(1-x_e^2)^2}
\left[
2x_e^4+x_e^2(1-x_e^2)+x_e^2x_{A'}^2
-\frac{1}{2}(1-x_e^2)\eta
\right],
\\
T_2
&=
\frac{16}{(1-x_e^2)\eta}
\left[
x_e^2(1-x_e^2)+x_e^2\eta+4x_e^4
\right.
\nonumber\\
&\hspace{2.9cm}
\left.
+x_{A'}^2(1-x_e^2-x_{A'}^2+\eta)
\right],
\\
T_3
&=
\frac{16}{\eta^2}
\left[
2x_e^4+x_e^2\eta+x_e^2x_{A'}^2
-\frac{1}{2}(1-x_e^2)\eta
\right].
\end{align}
where
\begin{equation}
\eta
=
x_{A'}^2
-\frac{1}{2}(1+x_e^2)(1-x_e^2+x_{A'}^2)
-\kappa(1-x_e^2)\cos\theta^\ast .
\end{equation}
Here $\theta^\ast$ is the COM-frame angle of the outgoing photon with
respect to the incoming dark-photon direction.

To obtain the lab-frame differential cross section with respect to the
outgoing photon energy, we boost from the COM frame to the lab frame.
The boost parameters are
\begin{equation}
\beta_{\rm cm}
=
\frac{p_{A'}^{\rm lab}}{E_{A'}+m_e},
\qquad
\gamma_{\rm cm}
=
\frac{E_{A'}+m_e}{\sqrt{s}},
\end{equation}
where
\begin{equation}
p_{A'}^{\rm lab}
=
\sqrt{E_{A'}^2-m_{A'}^2}.
\end{equation}
The lab-frame photon energy is
\begin{equation}
E_r
=
\gamma_{\rm cm}E_\gamma^\ast
\left(1+\beta_{\rm cm}\cos\theta^\ast\right),
\label{eq:Er_boost_appendix}
\end{equation}
so that
\begin{equation}
\cos\theta^\ast(E_r)
=
\frac{1}{\beta_{\rm cm}}
\left[
\frac{E_r}{\gamma_{\rm cm}E_\gamma^\ast}
-1
\right].
\label{eq:costheta_Er_appendix}
\end{equation}
The kinematically allowed lab-frame energy range is therefore
\begin{equation}
E_r^{\rm min}
=
\gamma_{\rm cm}E_\gamma^\ast(1-\beta_{\rm cm}),
\,
E_r^{\rm max}
=
\gamma_{\rm cm}E_\gamma^\ast(1+\beta_{\rm cm}).
\label{eq:Er_limits_appendix}
\end{equation}

Using
\begin{equation}
d\Omega^\ast = 2\pi\,d\cos\theta^\ast,
\qquad
\frac{d\cos\theta^\ast}{dE_r}
=
\frac{1}{\gamma_{\rm cm}\beta_{\rm cm}E_\gamma^\ast},
\end{equation}
we obtain
\begin{equation}
\frac{d\sigma_{A'e\to\gamma e}}{dE_r}
=
\frac{2\pi}{\gamma_{\rm cm}\beta_{\rm cm}E_\gamma^\ast}
\frac{d\sigma_{A'e\to\gamma e}}{d\Omega^\ast}
\Bigg|_{\cos\theta^\ast=\cos\theta^\ast(E_r)} .
\label{eq:inverse_dsigma_dEr_appendix}
\end{equation}
This expression is used in the event-rate calculation in
Sec.~\ref{sec:detection}.
\bibliographystyle{apsrev4-2}
\bibliography{refs}
\end{document}